%% file: main.tex
\documentclass[conference,hidelinks,a4paper]{IEEEtran}
\IEEEoverridecommandlockouts
\usepackage[utf8]{inputenc}

\usepackage[mathscr]{euscript} 
\usepackage[nolist]{acronym} 
\usepackage[normalem]{ulem} 
\usepackage[nospace,noadjust]{cite} 
\usepackage{algorithm}
\usepackage{algpseudocode}
\usepackage{amsmath,amssymb,amsfonts} 
\usepackage{amsthm} 
\usepackage{array}
\usepackage{float}
\usepackage{flushend} 
\usepackage{gensymb}
\usepackage{mathrsfs} 
\usepackage{multirow} 
\usepackage{soul} 
\usepackage{textcomp} 
\usepackage{xcolor}
\usepackage{color}
\usepackage{graphicx}
\usepackage{enumitem}
\usepackage[inkscapeformat=png]{svg}

\usepackage{tikz}
\tikzset{
  font={\fontsize{9pt}{12}\selectfont}}
\usetikzlibrary{plotmarks}
\usetikzlibrary{intersections,backgrounds, patterns}
\usetikzlibrary{patterns,shapes.arrows}

\usepackage{pgfplots}
\pgfplotsset{compat=newest} 
\usepgflibrary{arrows}
\usepgfplotslibrary{groupplots,dateplot}
\pgfplotsset{every axis/.append style={
	scaled x ticks = false,
	label style={font=\footnotesize},
	tick label style={font=\footnotesize},
	tick scale binop=\times}
}
\pgfkeys{/pgf/number format/.cd,
1000 sep={},
}

\usepackage[cmintegrals]{newtxmath} 
\usepackage[font=footnotesize]{subcaption}
\usepackage[font=footnotesize]{caption}
\usepackage[nolist]{acronym}
\usepackage{enumitem}
\DeclareUnicodeCharacter{2212}{−}

\newcommand{\fref}[1]{Fig.~\ref{#1}}

\newcommand{\tref}[1]{Table~\ref{#1}}
\newcommand{\eref}[1]{Eq.~\ref{#1}}

\definecolor{mycolor1}{RGB}{19, 133, 189}
\definecolor{mycolor2}{RGB}{230, 112, 32}
\definecolor{mycolor3}{RGB}{130, 173, 98}
\definecolor{mycolor4}{rgb}{0.49412,0.18431,0.55686}%
\definecolor{myhist}{RGB}{73, 80, 87}
\definecolor{hgreen}{rgb}{0, 0.5, 0}

\usepackage[left=14mm,right=14mm,top= 18mm,bottom=43mm]{geometry}

\def\BibTeX{{\rm B\kern-.05em{\sc i\kern-.025em b}\kern-.08em
    T\kern-.1667em\lower.7ex\hbox{E}\kern-.125emX}}

\begin{document}

\title{Computational Efficient Width-Wise Early Exiting in Wireless Communication Systems\thanks{The work of Hazem Sallouha was funded by the Research Foundation – Flanders (FWO), Postdoctoral Fellowship No. 12ZE222N.\\This work is supported by the 6G-Bricks project under the EU’s Horizon Europe Research and Innovation Program with Grant Agreement No. 101096954. }}


\author{\IEEEauthorblockN{Dieter Verbruggen, Hazem Sallouha, and Sofie Pollin}
\IEEEauthorblockA{\textit{Department of Electrical Engineering (ESAT) - WaveCoRE}\\
KU Leuven, 3000 Leuven, Belgium\\
E-mail:\{dieter.verbruggen, hazem.sallouha, sofie.pollin\}@kuleuven.be}}
\maketitle
\input{acronyms}
\input{0.abstract}


\input{1.Introduction}
\input{2.related_work}

\input{3.System_Model}

\input{4.architectures}
\input{5.Results}
\input{6.conclusions}

\bibliographystyle{ieeetr}
\bibliography{bibliography}

\end{document}

%% file: acronyms.tex
\begin{acronym}[HBCI]
\acro{amc}[AMC]{Automatic Modulation Classification}
\acro{6g}[6G]{6th Generation}
\acro{nr}[NR]{New Radio}
\acro{oran}[ORAN]{Open Radio Access Network}
\acro{rf}[RF]{Radio Frequency}
\acro{ru}[RU]{Radio Unit}
\acro{du}[DU]{Distributed Unit}
\acro{cu}[CU]{Central Unit}
\acro{iq}[IQ]{todododod}
\acro{cnn}[CNN]{Convolutional Neural Network}
\acro{ap}[AP]{access point}
\acro{ue}[UE]{user equipment units}
\acro{iq}[IQ]{In-phase/Quadrature}
\acro{fft}[FFT]{Fast Fourier Transform}
\acro{cp}[CP]{Cyclic Prefix}
\acro{bpsk}[BPSK]{binary phase shift keying}
\acro{qpsk}[QPSK]{quadrature phase shift keying}
\acro{qam}[QAM]{quadrature amplitude modulation}
\acro{egc}[EGC]{Equal Gain Combining}
\acro{snr}[SNR]{signal-to-noise ratio}
\acro{lb}[LB]{Likelihood-Based}
\acro{fb}[FB]{Feature-Based}
\acro{dl}[DL]{Deep Learning}
\acro{dnn}[DNN]{Deep Neural Networks}
\acro{oran}[O-RAN]{Open Radio Access Network}
\acro{flop}[FLOP]{floating-point operation}
\acro{mflop}[MFLOP]{Million FLOP}
\acro{awgn}[AWGN]{Additive White Gaussian Noise}
\acro{ee}[EE]{Early Exiting}
\end{acronym}

%% file: 0.abstract.tex
\begin{abstract}
Deep learning (DL) techniques are increasingly pervasive across various domains, including wireless communication, where they extract insights from raw radio signals. However, the computational demands of DL pose significant challenges, particularly in distributed wireless networks like Cell-free networks, where deploying DL models on edge devices becomes hard due to heightened computational loads. These computational loads escalate with larger input sizes, often correlating with improved model performance. To mitigate this challenge, Early Exiting (EE) techniques have been introduced in DL, primarily targeting the depth of the model. This approach enables models to exit during inference based on specified criteria, leveraging entropy measures at intermediate exits. Doing so makes less complex samples exit early, reducing the average computational load and inference time. In our contribution, we propose a novel width-wise exiting strategy for Convolutional Neural Network (CNN)-based architectures. 
By selectively adjusting the input size, we aim to regulate computational demands effectively. Our approach aims to decrease the average computational load during inference while maintaining performance levels comparable to conventional models. We specifically investigate Modulation Classification, a well-established application of DL in wireless communication. Our experimental results show substantial reductions in computational load, with an average decrease of 26\%,  and particularly notable reductions of 60\% in high-SNR scenarios. Through this work, we present a practical solution for reducing computational demands in deep learning applications, particularly within the domain of wireless communication.
\end{abstract}

%% file: 1.Introduction.tex
\input{acronyms}
\section{Introduction}

Recent advances in wireless communication show a shift towards distributed network paradigms, exemplified by emerging concepts like crowdsourced sensor networks \cite{rajendran2017electrosense} and cell-free networks \cite{chen2021survey}. Notably, in the development of \ac{6g} mobile networks, there is a trend towards cell-free distributed deployments \cite{distributeddeploy}, aimed at mitigating path-loss effects and ensuring homogeneous \ac{snr} across all user locations.

This transition towards distributed networks brings out a paradigm of distributed processing, which is essential for ensuring the scalability of such networks. Specifically, for \ac{dl}-based applications, this distributed approach has spurred the development of learning strategies such as federated learning \cite{MAL-083}, split learning \cite{split}, and distributed inference strategies \cite{verbruggen2024distributed,dist_inference}, where individual devices either make local decisions or collaboratively generate a central decision.

This trend towards distributed processing relocates computation to the network edge, catering to devices with energy constraints or computational load restrictions. In a study by \cite{EE_amc}, the authors explore the application of \ac{ee} techniques to \ac{dl}-based \ac{amc}. This concept involves making classification decisions at intermediate classifiers, allowing some input samples to exit early, thereby reducing computational load and speeding up inference time. The main idea behind \ac{ee} is that not all input samples require the same network depth to classify correctly.

Building upon the concept of \ac{ee}, our proposed method focuses on reducing computational load during inference by segmenting input samples and processing each segment sequentially using corresponding \textit{expert models}. At the classifier of each expert model, a decision is made to either accept the classification or delegate it to the next expert model. This width-wise implementation of early exiting enables us to tailor each expert model to target a specific subset of the dataset. Similar to the concept of \ac{ee}, which shows that not all input samples need the full depth of the network for correct classification, our width-wise approach showcases correct classification can be achieved with a segment of the input sample.

We implement our approach for \ac{amc}, a pivotal technology introduced approximately 15 years ago as a key technology for wireless networks \cite{haykin,Gardner} in the search towards more flexible and cognitive networks. Over the years, approaches in AMC have transitioned from statistical signal processing methods to \ac{dl}-based models \cite{amc_dl_first}, with the ResNet architecture \cite{oshea} widely recognized as one of the most performant feasible architectures. However, the high computational cost is often considered a barrier to implementing AMC in distributed and collaborative applications such as 6G Cell-free networks \cite{verbruggen2024distributed}. Our implementation aims to demonstrate width-wise \ac{ee}, lay the groundwork for further research in this area, and does not aim to surpass the state-of-the-art accuracy in \ac{amc}. 

\subsection{Contribution}
This paper proposes and evaluates a novel approach to reduce the average computational load of \ac{cnn}-based networks based on \ac{ee}. To the best of our knowledge, this work leads the way in describing and discussing a width-wise implementation of \ac{ee} as opposed to the commonly implemented depth-wise \ac{ee}. The reasoning behind width-wise \ac{ee} stems from the observation that the predominance computational load arises primarily from the initial convolutional layers before applying any pooling strategy. The computational load reduces significantly after each subsequent pooling operation as the features, which are the input of the subsequent convolutional layer, are reduced. 

Another challenge addressed in this paper is the selection of an adaptive exit criterion. Initially, exit criteria are established post-training and fine-tuned to balance model accuracy and time/computational load reduction. This iterative process involves testing various thresholds to determine optimal behavior. While such an implementation is feasible when dealing with only one intermediate exit, it becomes cumbersome when considering multiple exits. We propose and implement an adaptive exit criterion based on entropy, which enables the specification of targeted accuracy and the percentage of exits at each expert model. The selection of the exit criterion is data-driven and tailored to the current model.

The main contribution of this paper is twofold.
\begin{itemize}[leftmargin=*]
\item First, we propose a novel width-wise \ac{ee} approach based on ResNet and apply this approach to \ac{amc}. To analyze the reduction in computational load, we compare the proposed model against a similar baseline model comprising a direct implementation of a ResNet model. We show that our approach uses around 28\% less computations on average, peaking at 65\% less computations at high-SNR input samples while maintaining a similar accuracy. 
\item Second, we propose a dynamic method for selecting exit criteria based on a target accuracy and exit percentage at each expert model. By enabling the specification of these targets, the trade-off between model accuracy and exit percentage becomes more intuitive when training an EE model. This is particularly crucial in the context of research focused on EE models featuring multiple exits.
\end{itemize}

The remainder of the paper is structured as follows. Section II outlines the signal model and the AMC problem statement. Section III introduces the architectures of the baseline and proposed model. Subsequently, we present the performance evaluation results in Section IV. Finally, the paper concludes with final remarks in Section V.

%% file: 2.related_work.tex

%% file: 3.System_Model.tex
\section{Signal Model and Problem Statement}
We consider a scenario where a single transmitter transmits a modulated signal $s(t)$ through a flat fading channel $h(t)$ comprising attenuation and additive white noise $v(t)$. The modulation format $m$ employed by the transmitter is chosen from a set of known modulations $\mathcal{M}$. Upon reception, the continuous signal $r(t)$, denoted in \eref{eq:received}, is discretized into \ac{iq} samples. A collection of $N$ raw \ac{iq} samples forms a frame, represented as matrix \textbf{$R$} with dimensions $(N,2)$, where each column corresponds to the $I$ and $Q$ values. 
\begin{equation}
    r(t)=s(t)\times h(t) + v(t)\,.
    \label{eq:received}
\end{equation}
The objective of \ac{amc} is to correctly identify the transmitted modulation $\hat{m}$ from the known modulations $\mathcal{M}$ by analyzing the received signal according to a Maximum A Posteriori (MAP) criterion:
\begin{equation}
    \hat{m} = \arg \max_{m \in \mathcal{M}} \mathbb{P}(m | \textbf{R})
    \label{eq:s}
\end{equation}

%% file: 4.architectures.tex
\section{Architectures }

\begin{figure}
\begin{subfigure}{0.4\linewidth}
  \centering
    \includegraphics[]{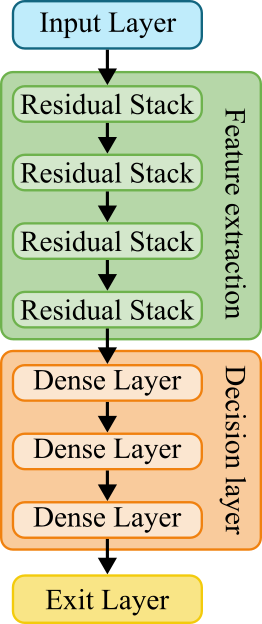}
    \caption{Baseline model, based on the ResNet model }
    \label{fig:baseline} 
\end{subfigure}
~
\begin{subfigure}{0.5\linewidth}
   \centering
    \includegraphics[]{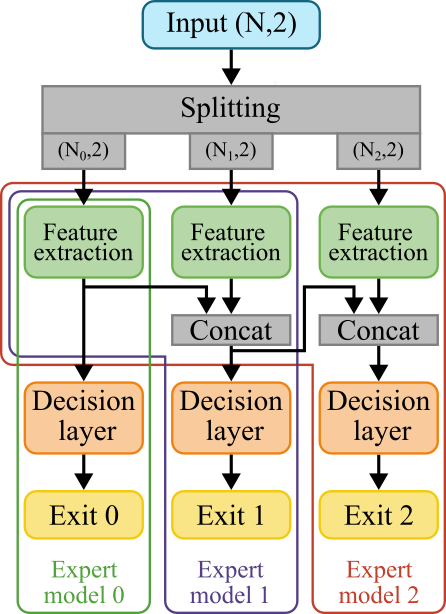}
    \caption{Proposed model containing 3 smaller expert models}
    \label{fig:proposedmodel} 
\end{subfigure}
\caption{Architectures of the baseline and the proposed model.}
\label{fig:models}
\end{figure}

In this paper, ResNets are the foundation of the baseline and proposed model. \fref{fig:baseline} offers a comprehensive overview of the ResNet model, systematically divided into four functional blocks for clarity in our discussion. Initially, we introduce the baseline model, providing detailed explanations of its four functional blocks. Subsequently, we delve into constructing the proposed model, covering its architectural development, training procedure, and inference method.

\subsection{Baseline model}
The baseline model is a direct implementation of a ResNet model. The first block, the input layer, interfaces real-world data with the model. The input of this block is the \ac{iq} samples derived from the dataset. The output of this block is a vector of dimensions $N\times 2$, with $N$ corresponding to the number of \ac{iq} samples in each frame.
The second block, dedicated to feature extraction, consists of multiple residual stacks. The received \ac{iq} samples are transformed into features using these residual stacks. The number of residual stacks is a design parameter, where each stack consists of 32 filters with a (3,1) kernel. 
Upon feature extraction, the third block, referred to as the decision layer, forms a soft decision. The soft decision comprises the probability assigned by the decision layer to identify the modulation utilized in the input IQ samples. The decision layer comprises two dense layers with 128 neurons each and a dense layer with six neurons (representing each modulation under consideration). 
Finally, the fourth and last block, the exit layer, concludes the model's prediction by selecting the modulation with the highest probability. This hierarchical organization of the ResNet model provides a structured framework for understanding its functionality and predictive capabilities.
These four blocks are used in addition to two non-trainable blocks to build the proposed model. 

\subsection{Proposed model}
The proposed model aims to reduce the average computational load while upholding the average classification accuracy. In traditional \ac{ee} techniques, computational load reduction involves truncating the model's depth and integrating exits in the depth dimension, allowing for layer skipping. However, for \ac{cnn}-based models, where pooling layers are utilized to decrease feature numbers, the heaviest computational load often resides in the initial layers of the network. The proposed model addresses this issue by focusing on these initial layers, as computational load diminishes with reduced input size, which is made possible due to the convolution operation performed by CNNs.

The architecture of the proposed model is illustrated in \fref{fig:proposedmodel}. Following the input layer of the proposed model, a splitting layer divides the original input vector of size $N$ into segments. Each segment corresponds to a distinct subset of the input vector: the first segment encompasses the initial $N_0$ values, the second segment comprises the subsequent $N_1$ values, and the final segment encompasses the remaining $N_2$ values, with $N_0+N_1+N_2=N$.

After the splitting layer, the model branches into three distinct sub-models. Each sub-model is tailored to address a specific subset of the dataset, classifying frames based on their difficulty level categorized as easy, medium, or hard. It is noteworthy that the model automatically conducts this categorization. Subsequent analysis reveals that the categorization closely aligns with factors such as the modulation type and SNR of the frames. Consequently, each sub-model specializes in classifying a particular subset of data, effectively becoming an \textit{expert model}. 

Each expert model comprises three main components: a feature extraction layer, a decision layer, and an exit layer. During feature extraction, the model utilizes a specified number of residual stacks, each consisting of 32 filters with a (3,1) kernel. Following feature extraction, the decision layer of each expert model consists of three dense layers. The first two dense layers each contains 128 neurons and the final dense layer contains six neurons.

In the case of subsequent expert models, features extracted from previous segments by the corresponding expert are also considered. This is achieved by concatenating these features with the features extracted from the current segment, ensuring that the model integrates information from previous segments seamlessly.

\subsection{Proposed Model Training}
Training the proposed model unfolds into two phases. In the first phase, the joint phase, the whole model is jointly trained on all the training frames of the dataset. To this end, we adapt the loss function proposed in \cite{branchynet} with all weights set to one. This loss function is, therefore, the sum of the categorical loss at each exit of the model given by \eref{eq:loss}. 

\begin{equation}
    L_{joint}(\hat{\textbf{y}},\textbf{y};\theta)=\sum_{e=0}^{2} L(\hat{\textbf{y}}_{e},\textbf{y};\theta) 
    \label{eq:loss}
\end{equation}
\noindent\makebox[\linewidth]{\rule{\textwidth}{0pt}} 

Each expert model is individually and sequentially retrained in the second training phase, starting with expert model 0. The second training phase consists of two steps. In the first step, one expert model is retrained using (a subset of) the training frames of the dataset. Upon retraining, the weights of this expert model are frozen, rendering them no longer trainable.

The second training phase's second step determines an exit criterion using the training frames. Once an exit criterion is selected, the training frames are divided into two subsets: frames that would exit at this point and frames that would not. Frames that exit are excluded from further training steps, while the remaining frames are used to retrain the subsequent expert model. No exit criterion is selected for the last expert model, as all remaining frames exit at this stage.

By the end of the second training phase, the three expert models specialize in extracting distinct subsets of the dataset. Expert model 0 targets easy frames, expert model 1 handles medium frames, and the final expert model processes all frames that have not exited yet. Consequently, the dataset is divided into three subsets accordingly. After the second training phase, the model is completely trained, and the exit criterion at each exit is dynamically determined.

\subsection{Proposed Model Inference}
During inference, the objective is to classify the modulation of an unseen frame. Initially, the frame is segmented into segments corresponding to each expert model. Expert model 0 initiates the process by processing the first segment. Expert model 0 decides whether to classify the frame immediately or delegate the classification to the next expert model based on the entropy, calculated from the soft decision, and the exit threshold, dynamically learned during training. 

If expert model 0 delegates the classification, expert model 1 begins processing its corresponding segment of the unseen frame. Upon completing the feature extraction layer, the extracted features from expert model 0 are concatenated with those from expert model 1. Similarly, at the decision layer, expert model 1 determines whether to classify the frame or delegate the classification to the final expert model based on the exit criterion obtained.

Should expert model 1 delegate the classification, the final expert model 2 performs the classification based on the features extracted from the final segment and those extracted by the two preceding expert models. The final expert model 2 cannot delegate the classification further and proceeds to classify the frame. When an expert model decides to classify, the objective is achieved, and no further processing is done. This results in a lower computational load when classifying at an earlier than a later exit.

\subsection{Exit Criterion Selection}
\begin{figure}
    \centering
    \input{figures/tex_fig/threshold}
    \caption{Exit threshold selection by averaging the entropy where the accuracy drops below 95\% and the entropy for which at least 25\% of the training frames exit.}
    \label{fig:th}
    \vspace{-0.5cm}
\end{figure}
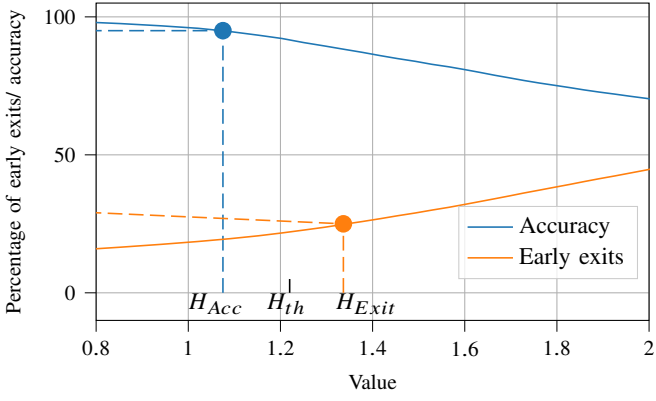

At each exit, except for the final one, we derive an exit criterion based on the cross-categorical entropy of the soft decision (\textbf{Z}) as defined in \eref{eq:entropy}. Here, $z_i$ represents the confidence level of each classification class, which is equal to the output of the decision layer.

\begin{equation}
H(\textbf{Z}) = \sum_{i=0}^{5}-z_ilog_2(z_i).
\label{eq:entropy}
\end{equation}
Selecting an appropriate threshold entropy results in the trade-off between classification accuracy and computational load reduction. \fref{fig:th} illustrates the relationship between accuracy and the percentage of frames exiting based on different threshold entropy values. It is important to note that the classification accuracy is calculated only for the subset of data that exits at the specified threshold.
The figure shows that accuracy decreases as the threshold entropy increases while the percentage of exiting frames rises. To determine the threshold, we employ two key considerations as follows.
\begin{itemize}[leftmargin=*]
\item We first identify the entropy $H_{Acc}$ at which the accuracy drops below a predetermined percentage. We opt for a threshold of 95\%, ensuring that frames exiting at this point are classified with a sufficiently high accuracy. Lowering this threshold may result in a notable decrease in overall model accuracy.
\item Next, we select the threshold $H_{exits}$ at which at least 25\% of the training frames are exiting. Choosing a lower $H_{exits}$ value may underutilize the exit, resulting in a less substantial reduction in computational load.
\end{itemize}
    
Finally, we compute the average of these two selected values to establish the exit criterion $H_{th}$. The selection of the desired accuracy (95\%) and exit percentage (25\%) is based on empirical analysis, presenting a trade-off between achieving high accuracy and optimizing exit efficiency, which results in a lower computational load. 

%% file: figures/tex_fig/threshold.tex
\begin{tikzpicture}

\definecolor{darkgray176}{RGB}{176,176,176}
\definecolor{darkorange25512714}{RGB}{255,127,14}
\definecolor{lightgray204}{RGB}{204,204,204}
\definecolor{steelblue31119180}{RGB}{31,119,180}

\begin{axis}[
legend cell align={left},
legend style={
  fill opacity=0.8,
  draw opacity=1,
  text opacity=1,
  at={(0.98,0.13)},
  anchor=south east,
  draw=lightgray204
},
tick align=outside,
tick pos=left,
x grid style={darkgray176},
xlabel={Value},
xmajorgrids,
xmin=0.8, xmax=2,
xtick style={color=black},
y grid style={darkgray176},
ylabel={Percentage of early exits/ accuracy},
ymajorgrids,
ymin=-10, ymax=105,
ytick style={color=black},
width=\linewidth,
height=0.65\linewidth,
]
\addplot [semithick, steelblue31119180]
table {%
0.00149733755817073 100
0.0346366636360578 100
0.0509500054284687 100
0.0656668080878599 99.9666777740753
0.080669580274431 99.9750062484379
0.0985977397309736 99.9800039992002
0.123985109222907 99.9833361106482
0.156188731609979 99.9428653049564
0.197966972135886 99.9000124984377
0.260465574948588 99.8111320964337
0.334481941880897 99.7300269973003
0.408610269518918 99.5364057812926
0.488533144118965 99.3750520789934
0.571452601431984 99.161602953619
0.663307030409615 98.7358045853868
0.757461727342536 98.2601159922672
0.853084511178885 97.5626523342291
0.941590050983252 96.7472501617552
1.02156602522675 95.8557857896784
1.09176005624581 94.6529130045787
1.15235728336406 93.3503324833758
1.2035977083708 92.1527546307319
1.25059454745883 90.6822417162856
1.29621260163489 89.4395895830616
1.33658845262999 88.2879880005
1.37787933731902 87.1365145394184
1.41680494461073 85.9505403638322
1.45338462255334 84.9042628050813
1.49286513186806 83.8827184743402
1.5284626168916 82.8109375538775
1.5661469394007 81.8806039798673
1.59989045144794 80.8909390019677
1.63274859680853 79.8818786912909
1.66551834308658 78.8218538832157
1.69763644710051 77.8771212611394
1.72839777998179 76.9463729607725
1.76172010651706 76.0256659537235
1.79431116282086 75.2141834004486
1.82754199458588 74.3743585695113
1.85885814349884 73.5340119484116
1.89148376831546 72.7531811704707
1.9261351767766 71.993366015463
1.95886209937459 71.2316373419681
1.99163683314951 70.5169647217507
2.02276156537206 69.7802322674485
2.05754574443967 69.0895757872047
2.09666577811818 68.3811221495185
2.13621955204216 67.5900512755048
2.17387240172616 66.8298577112977
2.21197842569518 66.1578335136018
2.24826332537422 65.4726905461891
2.27904220048588 64.8026509284132
2.30739698013363 64.1795350089421
2.33288704476596 63.5308767759099
2.35667379322414 62.8914279365197
2.3774950448335 62.2552317230596
2.39831526454695 61.6221138908234
2.41440080188215 61.0305082366976
2.43087226828589 60.4748193996655
2.44411734015943 59.9193234012983
2.45624452480237 59.2923451275812
2.4671036813605 58.6908411337519
2.47641465009556 58.1393848486315
2.48525696610703 57.5705147537341
2.4932357572285 57.0381712785738
2.50048637380623 56.537591729358
2.50701277144117 56.0339994848563
2.5129189996371 55.5305144699333
2.51824392647312 55.0580138527375
2.52334441594975 54.5687743655889
2.52802423586238 54.0949415008357
2.53200735405636 53.6485401614062
2.53587833956198 53.2103720781656
2.53949152307396 52.7746195257599
2.54278378695943 52.3357792462264
2.54578368033843 51.9046412714497
2.54861692252874 51.4414284022579
2.55137122468231 51.0616745237075
2.55382576107542 50.669863206882
2.55627960953793 50.2791103910077
2.55840239139753 49.8931263359208
2.56050212758192 49.5351909235688
2.56238877726188 49.1420836331264
2.56416810976771 48.7596534981506
2.56594709830823 48.4148998226212
2.56755788781622 48.0817872730909
2.5690699592826 47.7238636760038
2.57051134587948 47.3891104699946
2.57191455232787 47.0608288542176
2.57320614203585 46.7489129335625
2.57448896062864 46.4250397217809
2.57568871160106 46.1291634157866
2.5768442630322 45.8092846816882
2.57802956746195 45.5124138450124
2.5792155598223 45.2165402495718
2.58055582062047 44.9258428858644
2.58207924294182 44.6182852261956
};
\addlegendentry{Accuracy}
\path [draw=steelblue31119180, semithick, dash pattern=on 5.55pt off 2.4pt]
(axis cs:0,95)
--(axis cs:1.07519864291567,95);

\path [draw=steelblue31119180, semithick, dash pattern=on 5.55pt off 2.4pt]
(axis cs:1.07519864291567,0)
--(axis cs:1.07519864291567,95);

\addplot [semithick, steelblue31119180, mark=*, mark size=3, mark options={solid}, only marks, forget plot]
table {%
1.07519864291567 95
};
\addplot [semithick, darkorange25512714]
table {%
0.00149733755817073 0.00103339947089947
0.0346366636360578 1.03443287037037
0.0509500054284687 2.06783234126984
0.0656668080878599 3.10123181216931
0.080669580274431 4.13463128306878
0.0985977397309736 5.16803075396825
0.123985109222907 6.20143022486772
0.156188731609979 7.2348296957672
0.197966972135886 8.26822916666667
0.260465574948588 9.30162863756614
0.334481941880897 10.3350281084656
0.408610269518918 11.3684275793651
0.488533144118965 12.4018270502646
0.571452601431984 13.435226521164
0.663307030409615 14.4686259920635
0.757461727342536 15.502025462963
0.853084511178885 16.5354249338624
0.941590050983252 17.5688244047619
1.02156602522675 18.6022238756614
1.09176005624581 19.6356233465608
1.15235728336406 20.6690228174603
1.2035977083708 21.7024222883598
1.25059454745883 22.7358217592593
1.29621260163489 23.7692212301587
1.33658845262999 24.8026207010582
1.37787933731902 25.8360201719577
1.41680494461073 26.8694196428571
1.45338462255334 27.9028191137566
1.49286513186806 28.9362185846561
1.5284626168916 29.9696180555556
1.5661469394007 31.003017526455
1.59989045144794 32.0364169973545
1.63274859680853 33.069816468254
1.66551834308658 34.1032159391534
1.69763644710051 35.1366154100529
1.72839777998179 36.1700148809524
1.76172010651706 37.2034143518519
1.79431116282086 38.2368138227513
1.82754199458588 39.2702132936508
1.85885814349884 40.3036127645503
1.89148376831546 41.3370122354497
1.9261351767766 42.3704117063492
1.95886209937459 43.4038111772487
1.99163683314951 44.4372106481481
2.02276156537206 45.4706101190476
2.05754574443967 46.5040095899471
2.09666577811818 47.5374090608466
2.13621955204216 48.570808531746
2.17387240172616 49.6042080026455
2.21197842569518 50.637607473545
2.24826332537422 51.6710069444444
2.27904220048588 52.7044064153439
2.30739698013363 53.7378058862434
2.33288704476596 54.7712053571429
2.35667379322414 55.8046048280423
2.3774950448335 56.8380042989418
2.39831526454695 57.8714037698413
2.41440080188215 58.9048032407407
2.43087226828589 59.9382027116402
2.44411734015943 60.9716021825397
2.45624452480237 62.0050016534392
2.4671036813605 63.0384011243386
2.47641465009556 64.0718005952381
2.48525696610703 65.1052000661376
2.4932357572285 66.138599537037
2.50048637380623 67.1719990079365
2.50701277144117 68.205398478836
2.5129189996371 69.2387979497355
2.51824392647312 70.2721974206349
2.52334441594975 71.3055968915344
2.52802423586238 72.3389963624339
2.53200735405636 73.3723958333333
2.53587833956198 74.4057953042328
2.53949152307396 75.4391947751323
2.54278378695943 76.4725942460317
2.54578368033843 77.5059937169312
2.54861692252874 78.5393931878307
2.55137122468231 79.5727926587302
2.55382576107542 80.6061921296296
2.55627960953793 81.6395916005291
2.55840239139753 82.6729910714286
2.56050212758192 83.7063905423281
2.56238877726188 84.7397900132275
2.56416810976771 85.773189484127
2.56594709830823 86.8065889550265
2.56755788781622 87.8399884259259
2.5690699592826 88.8733878968254
2.57051134587948 89.9067873677249
2.57191455232787 90.9401868386243
2.57320614203585 91.9735863095238
2.57448896062864 93.0069857804233
2.57568871160106 94.0403852513228
2.5768442630322 95.0737847222222
2.57802956746195 96.1071841931217
2.5792155598223 97.1405836640212
2.58055582062047 98.1739831349206
2.58207924294182 99.2073826058201
};
\addlegendentry{Early exits}
\path [draw=darkorange25512714, semithick, dash pattern=on 5.55pt off 2.4pt]
(axis cs:0,35)
--(axis cs:1.33658845262999,25);

\path [draw=darkorange25512714, semithick, dash pattern=on 5.55pt off 2.4pt]
(axis cs:1.33658845262999,0)
--(axis cs:1.33658845262999,25);

\addplot [semithick, darkorange25512714, mark=*, mark size=3, mark options={solid}, only marks, forget plot]
table {%
1.33658845262999 25
};
\path [draw=black, semithick, dash pattern=on 5.55pt off 2.4pt]
(axis cs:1.22,0)
--(axis cs:1.22,5);

\draw (axis cs:0.98,-6) node[
  scale=1,
  anchor=base west,
  text=black,
  rotate=0.0
]{$H_{Acc}$};
\draw (axis cs:1.3,-6) node[
  scale=1,
  anchor=base west,
  text=black,
  rotate=0.0
]{$H_{Exit}$};
\draw (axis cs:1.15,-6) node[
  scale=1,
  anchor=base west,
  text=black,
  rotate=0.0
]{$H_{th}$};
\end{axis}

\end{tikzpicture}

%% file: 5.Results.tex
\section{Results and discussion}
To mitigate the impact of randomness inherent in deep learning when dealing with a limited-size dataset, we conduct 32 Monte Carlo simulations and average the results for both the baseline and the proposed model. The baseline model undergoes training for 40 epochs in each simulation, employing early stopping based on validation frames. Similarly, the proposed model undergoes training with 40 epochs per training step and early stopping. 

We quantify the performance of both the proposed model and the baseline model using two key metrics: model accuracy and computational load, which are defined as follows.
\begin{itemize}[leftmargin=*]
\item \textit{Model accuracy} represents the percentage of correctly classified modulation schemes. A higher accuracy value indicates better performance.

\item \textit{Computational load} is evaluated by the number of \acp{flop}required for a model to perform one modulation classification. A lower \ac{flop} count indicates a reduced computational burden. We estimate \acp{flop} using TensorFlow's built-in estimator. However, it is important to note that these estimates serve as approximations and may vary depending on hardware configurations. 

The \acp{flop} offers insights into metrics like power consumption and inference time. However, since these metrics are hardware-dependent, we do not address them in this paper.
\end{itemize}

In the subsequent sections, we will initially present the dataset created for the results, followed by an overview of the model design parameters and, ultimately, a comparison between the proposed and baseline models.

\subsection{Dataset}
A synthetic dataset is generated based upon the recently released RML22 \cite{RadioML22}, using Python and gnuradio \cite{gnuradio}. Our dataset considers six different modulation techniques $\mathcal{M} =\{$BPSK , QPSK, PSK8, PAM4, QAM16, QAM64 $\}$, which are a subsection of the modulation techniques considered in RML22.
We concentrate on similar digital modulations, making the classification task more challenging. As a result, the accuracy reported here should not be directly compared to studies using the full RML22 dataset, as the different dataset composition affects the classification challenge. The SNR of the signal chosen ranges from $-20dB$ to $20dB$ with a step of $2dB$, resulting in 21 different \acp{snr}. 
For each pair of modulation and SNR, we generated 1024 frames of 1024 \ac{iq} samples. To ensure an equal representation of each SNR/Modulation pair, the dataset is split by randomly picking 768 frames for training, 128 frames for validation, and the remaining 128 frames for testing for each SNR/modulation pair. This resulted in a total of 112,880 frames, of which 98,304 were used for training, 12,288 for validation, and the final 12,288 for testing.

\subsection{Model Design Parameters}
To maintain the comparability between the baseline and proposed models, both models utilize an input size of (512,2), representing a vector of IQ samples. We adopt a configuration consisting of 4 residual stacks for the baseline model's feature extraction. Similarly, each feature extraction layer in the proposed model, mirroring the baseline structure, encompasses four residual stacks, effectively partitioning the baseline's feature layer into three segments. The total computational workload for feature extraction in the baseline model matches the cumulative workload across all feature extraction layers in the proposed model.

While the proposed model introduces additional design parameters, we uphold consistency in the input vector size to facilitate a fair comparison. This input vector is split into three segments: the first segment comprises 32 IQ samples, 160 samples for the second segment, and the remaining 320 samples for the final segment. This split is chosen as it indicates the potential of width-wise early exiting. The consideration of different splits is beyond the scope of this paper.

\subsection{Computational load}
The computational load of the baseline model is fixed, while for the proposed model, the computational load depends on which exit has been taken. The computational load for the baseline and each expert model are shown in \tref{tab:flops}. 
The selection of a small segment size for expert model 0 results in a computational requirement that is an order of magnitude lower than the baseline and the other expert models. Expert model 2, on the other hand, exhibits a slightly higher computational load compared to the baseline model. This is due to the intermediate checks performed by the expert models 0 and expert model 1. We also provide the amount of trainable parameters, which is usually linked with the model complexity. However, the number of parameters is not a good metric when comparing the computational load. Both the baseline model and Exit 1 have similar trainable parameters; however, the computational load of the baseline model is threefold.

\begin{table}[]
\caption{MFLOPs and amount of trainable parameters for baseline model and each exit of the proposed model}
    \label{tab:flops}
    \centering
\begin{tabular}{|c|c|c|}
\cline{2-3} 
    \multicolumn{1}{c|}{ }  & \multicolumn{1}{c|}{total MFLOPs}  & \multicolumn{1}{c|}{params} \\ \cline{2-3} \hline

    \multicolumn{1}{|c|}{Baseline model }  & \multicolumn{1}{c|}{12.53}  & \multicolumn{1}{c|}{202 438} \\ \hline \hline
    \multicolumn{1}{|c|}{Expert model 0 }  & \multicolumn{1}{c|}{0.80} & \multicolumn{1}{c|}{79558} \\ \hline
    \multicolumn{1}{|c|}{Expert model 1 }  & \multicolumn{1}{c|}{4.73} & \multicolumn{1}{c|}{200076} \\ \hline
    \multicolumn{1}{|c|}{Expert model 2 }  & \multicolumn{1}{c|}{12.62} & \multicolumn{1}{c|}{402514} \\ \hline
    
\end{tabular}
\end{table}

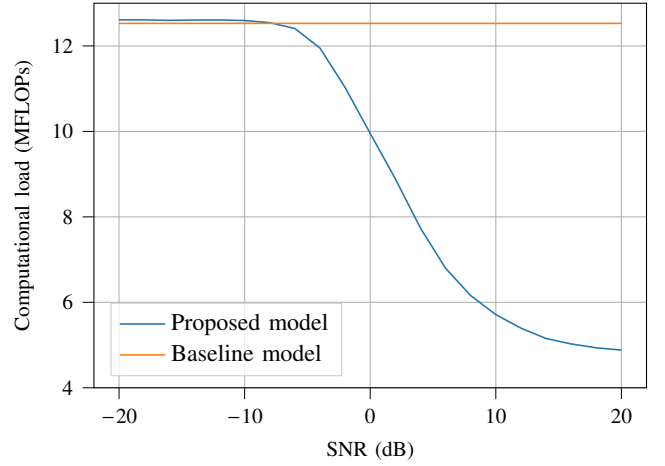
\begin{figure}
    \centering
    \input{figures/tex_fig/reduction}
    \caption{Computational load of the proposed and baseline model.}
    \label{fig:reduction}
    \vspace{-0.3cm}
\end{figure}

The average computational load is calculated by multiplying the computational load of each expert model by the percentage of frames exiting at the corresponding expert model. \fref{fig:reduction} illustrates the reduction in computational load achieved by the proposed model compared to the baseline model. For low-SNR scenarios, there is a minor increase in computational cost attributable to the intermediate checks.

Conversely, for higher SNR values, the reduction in computational load is significant, reaching up to 60\%. This graph demonstrates that the proposed model prioritizes the exiting of frames with higher SNR, generally regarded as easier samples. On average, across all SNR levels, the proposed model achieves a 26.2\% reduction in computational load.

\subsection{Classification accuracy}

\begin{figure}
    \centering
    \input{figures/tex_fig/accuracy}
    \caption{Reduction in computational load of the proposed model compared to the baseline model.}
    \label{fig:accuracy}
    \vspace{-0.3cm}
\end{figure}
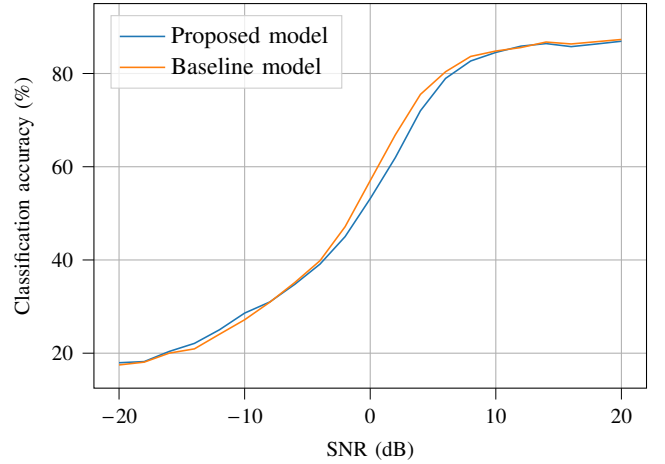

The classification accuracies for both the baseline and proposed models are shown in \fref{fig:accuracy}. As expected, the accuracies follow a typical pattern: lower accuracies at lower SNR levels, gradually increasing as SNR values rise. The accuracy reaches a peak of approximately 87\% for SNRs above 10 dB.

Comparing the proposed model with the baseline reveals similar behavior for both low-SNR and high-SNR samples. However, in the mid-SNR range, the baseline model outperforms the proposed model in terms of accuracy. \tref{tab:accuracy} presents the average accuracy for different SNR ranges, showing that overall, the baseline model outperforms the proposed model by 0.7\%.

\begin{table}[htbp]
\centering
\caption{Average accuracy for the baseline model and proposed model for different SNR ranges}
\label{tab:accuracy}
\begin{tabular}{l|c|c|c|c|}
\cline{1-5}
 
\multicolumn{1}{|l|}{SNR range (dB) }   & [-20,20] &[-20,-7] & [-7,7] & [7,20] \\ \hline \hline                     
\multicolumn{1}{|l|}{Baseline model }  & 55.32\% &22.67\% &57.41\% &85.89\%\\ \hline
\multicolumn{1}{|l|}{Proposed model }  & 54.61\% & 23.32\% & 55.00\% &85.50\%\\ \hline
\end{tabular}
\end{table}

\subsection{Early exiting frequency}
\begin{figure}
    \centering
    \input{figures/tex_fig/exits_snr}
    \caption{Percentage of frames exiting at each exit in function of the SNR.}
    \label{fig:exits}
    \vspace{-0.2cm}
\end{figure}
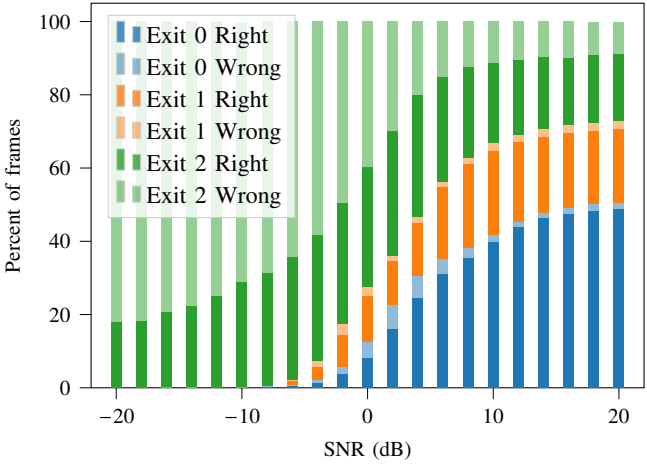

In \fref{fig:exits}, the distribution of frame exits across various SNR levels is illustrated, with each exit represented by a distinct color. The figure also differentiates between correctly classified and misclassified frames at each exit. A clear trend emerges, showing a correlation between frame SNR and exit selection: higher SNR frames tend to exit at earlier stages, while the final expert model typically classifies lower SNR frames. Additionally, the figure categorizes the dataset into 'easy,' 'medium,' and 'hard' subsets based on classification difficulty.
\tref{tab:exits} expands on the percentage of frames exiting at each expert model, demonstrating that the exit percentage is influenced not only by SNR but also by modulation type. For instance, modulations like BPSK and PAM4 tend to exit more frequently at expert model 0, while QAM16 and QAM64 are more likely to exit at expert model 2. This indicates that the efficiency of width-wise EE depends on the characteristics of the dataset.
In addition to the exit percentages per modulation, \tref{tab:exits} also provides the exit percentages for the entire dataset, which can help in selecting suitable exit threshold criteria. For the dataset used in this section, 32.5\% of samples exit early, at either the first or second exit.

\begin{table}[]
\caption{The percentage of frames exiting at each exit and the corresponding classification accuracy of the exited frames.}
    \label{tab:exits}
    \centering
\begin{tabular}{|l|c|c|c|}
\cline{2-4}
    \multicolumn{1}{c|}{ }& Expert model 0  &Expert model 1& Expert model 2\\ \hline
    
    BPSK  &  47.6\% & 5.9\%  & 46.4\% \\ \hline
    QPSK  &  21.1\% & 16.4\% & 62.5\% \\ \hline
    PSK8  &  3.9\%  & 28.2\% & 67.8\% \\ \hline
    PAM4  &  49.4\% & 6.9\%  & 43.7\% \\ \hline 
    QAM16 &  0.9\%  & 5.7\%  & 93.4\% \\ \hline 
    QAM64 &  1.0\%  & 7.6\%  & 91.4\% \\ \hline \hline
    Mean  &  20.7\% & 11.8\% & 67.5\% \\ \hline  
    
\end{tabular}
\end{table}

%% file: figures/tex_fig/reduction.tex
\begin{tikzpicture}

\definecolor{darkgray176}{RGB}{176,176,176}
\definecolor{darkorange25512714}{RGB}{255,127,14}
\definecolor{lightgray204}{RGB}{204,204,204}
\definecolor{steelblue31119180}{RGB}{31,119,180}

\begin{axis}[
legend cell align={left},
legend style={
  fill opacity=0.8,
  draw opacity=1,
  text opacity=1,
  at={(0.03,0.03)},
  anchor=south west,
  draw=lightgray204
},
tick align=outside,
tick pos=left,
x grid style={darkgray176},
xlabel={SNR (dB)},
xmajorgrids,
xmin=-22, xmax=22,
xtick style={color=black},
y grid style={darkgray176},
ylabel={Computational load (MFLOPs)},
ymajorgrids,
ymin=4, ymax=13,
ytick style={color=black},
width=\linewidth,
height=0.75\linewidth,
]

\addplot [semithick, steelblue31119180]
table {%
-20 12.6106887965902
-18 12.6101283565267
-16 12.5989848254395
-14 12.6061195479329
-12 12.6074812461344
-10 12.5948114847819
-8 12.5443757753499
-6 12.4086761848958
-4 11.9492747644043
-2 11.0282174542236
0 9.95236337300618
2 8.88933787638346
4 7.73686815795898
6 6.79783352722168
8 6.15799877624512
10 5.71273192993164
12 5.39540455607096
14 5.15382125528971
16 5.02581041320801
18 4.93587947591146
20 4.88199728637695
};
\addlegendentry{Proposed model}
\addplot [semithick, darkorange25512714]
table {%
-20 12.52853
-18 12.52853
-16 12.52853
-14 12.52853
-12 12.52853
-10 12.52853
-8 12.52853
-6 12.52853
-4 12.52853
-2 12.52853
0 12.52853
2 12.52853
4 12.52853
6 12.52853
8 12.52853
10 12.52853
12 12.52853
14 12.52853
16 12.52853
18 12.52853
20 12.52853
};
\addlegendentry{Baseline model}
\end{axis}


\end{tikzpicture}

%% file: figures/tex_fig/accuracy.tex
\begin{tikzpicture}

\definecolor{darkgray176}{RGB}{176,176,176}
\definecolor{darkorange25512714}{RGB}{255,127,14}
\definecolor{lightgray204}{RGB}{204,204,204}
\definecolor{steelblue31119180}{RGB}{31,119,180}

\begin{axis}[
legend cell align={left},
legend style={
  fill opacity=0.8,
  draw opacity=1,
  text opacity=1,
  at={(0.03,0.97)},
  anchor=north west,
  draw=lightgray204
},
tick align=outside,
tick pos=left,
x grid style={darkgray176},
xlabel={SNR (dB)},
xmajorgrids,
xmin=-22, xmax=22,
xtick style={color=black},
y grid style={darkgray176},
ylabel={Classification accuracy (\%)},
ymajorgrids,
ymin=12.5, ymax=95,
ytick style={color=black},
width=\linewidth,
height=0.75\linewidth,
]
\addplot [semithick, steelblue31119180]
table {%
-20 17.9524739583333
-18 18.1966145833333
-16 20.3816731770833
-14 22.100830078125
-12 25.0325520833333
-10 28.6031087239583
-8 30.9855143229167
-6 34.7900390625
-4 39.1072591145833
-2 45.00732421875
0 53.1595865885417
2 61.962890625
4 72.0174153645833
6 78.948974609375
8 82.6863606770833
10 84.521484375
12 85.8683268229167
14 86.4176432291667
16 85.7645670572917
18 86.3260904947916
20 86.9222005208333
};
\addlegendentry{Proposed model}
\addplot [semithick, darkorange25512714]
table {%
-20 17.4784342447917
-18 18.07861328125
-16 20.0154622395833
-14 20.928955078125
-12 24.072265625
-10 27.1728515625
-8 30.9285481770833
-6 35.174560546875
-4 39.8356119791667
-2 47.137451171875
0 57.0210774739583
2 66.845703125
4 75.5228678385417
6 80.3792317708333
8 83.6649576822916
10 84.8225911458333
12 85.5712890625
14 86.7513020833333
16 86.3382975260417
18 86.8204752604167
20 87.298583984375
};
\addlegendentry{Baseline model}
\end{axis}

\end{tikzpicture}

%% file: figures/tex_fig/exits_snr.tex
\begin{tikzpicture}

\definecolor{darkgray176}{RGB}{176,176,176}
\definecolor{darkorange25512714}{RGB}{255,127,14}
\definecolor{forestgreen4416044}{RGB}{44,160,44}
\definecolor{lightgray204}{RGB}{204,204,204}
\definecolor{steelblue31119180}{RGB}{31,119,180}

\begin{axis}[
legend cell align={left},
legend style={
  fill opacity=0.8,
  draw opacity=1,
  text opacity=1,
  at={(0.03,0.97)},
  anchor=north west,
  draw=lightgray204
},
tick align=outside,
tick pos=left,
x grid style={darkgray176},
xlabel={SNR (dB)},
xmin=-22, xmax=22,
xtick style={color=black},
y grid style={darkgray176},
ylabel={Percent of frames},
ymin=0, ymax=105,
ytick style={color=black},
width=\linewidth,
height=0.75\linewidth,
]
\draw[draw=none,fill=steelblue31119180] (axis cs:-20.4,0) rectangle (axis cs:-19.6,0.0142415364583333);
\addlegendimage{ybar,ybar legend,draw=none,fill=steelblue31119180}
\addlegendentry{Exit 0 Right}

\draw[draw=none,fill=steelblue31119180] (axis cs:-18.4,0) rectangle (axis cs:-17.6,0.0162760416666667);
\draw[draw=none,fill=steelblue31119180] (axis cs:-16.4,0) rectangle (axis cs:-15.6,0.0874837239583333);
\draw[draw=none,fill=steelblue31119180] (axis cs:-14.4,0) rectangle (axis cs:-13.6,0.0406901041666667);
\draw[draw=none,fill=steelblue31119180] (axis cs:-12.4,0) rectangle (axis cs:-11.6,0.030517578125);
\draw[draw=none,fill=steelblue31119180] (axis cs:-10.4,0) rectangle (axis cs:-9.6,0.0528971354166667);
\draw[draw=none,fill=steelblue31119180] (axis cs:-8.4,0) rectangle (axis cs:-7.6,0.223795572916667);
\draw[draw=none,fill=steelblue31119180] (axis cs:-6.4,0) rectangle (axis cs:-5.6,0.510660807291667);
\draw[draw=none,fill=steelblue31119180] (axis cs:-4.4,0) rectangle (axis cs:-3.6,1.18611653645833);
\draw[draw=none,fill=steelblue31119180] (axis cs:-2.4,0) rectangle (axis cs:-1.6,3.564453125);
\draw[draw=none,fill=steelblue31119180] (axis cs:-0.4,0) rectangle (axis cs:0.4,8.09733072916667);
\draw[draw=none,fill=steelblue31119180] (axis cs:1.6,0) rectangle (axis cs:2.4,16.0176595052083);
\draw[draw=none,fill=steelblue31119180] (axis cs:3.6,0) rectangle (axis cs:4.4,24.4405110677083);
\draw[draw=none,fill=steelblue31119180] (axis cs:5.6,0) rectangle (axis cs:6.4,30.9163411458333);
\draw[draw=none,fill=steelblue31119180] (axis cs:7.6,0) rectangle (axis cs:8.4,35.2803548177083);
\draw[draw=none,fill=steelblue31119180] (axis cs:9.6,0) rectangle (axis cs:10.4,39.8152669270833);
\draw[draw=none,fill=steelblue31119180] (axis cs:11.6,0) rectangle (axis cs:12.4,43.7744140625);
\draw[draw=none,fill=steelblue31119180] (axis cs:13.6,0) rectangle (axis cs:14.4,46.1527506510417);
\draw[draw=none,fill=steelblue31119180] (axis cs:15.6,0) rectangle (axis cs:16.4,47.3002115885417);
\draw[draw=none,fill=steelblue31119180] (axis cs:17.6,0) rectangle (axis cs:18.4,48.1465657552083);
\draw[draw=none,fill=steelblue31119180] (axis cs:19.6,0) rectangle (axis cs:20.4,48.7223307291667);
\draw[draw=none,fill=steelblue31119180,fill opacity=0.5] (axis cs:-20.4,0.0142415364583333) rectangle (axis cs:-19.6,0.0732421875);
\addlegendimage{ybar,ybar legend,draw=none,fill=steelblue31119180,fill opacity=0.5}
\addlegendentry{Exit 0 Wrong}

\draw[draw=none,fill=steelblue31119180,fill opacity=0.5] (axis cs:-18.4,0.0162760416666667) rectangle (axis cs:-17.6,0.0874837239583333);
\draw[draw=none,fill=steelblue31119180,fill opacity=0.5] (axis cs:-16.4,0.0874837239583333) rectangle (axis cs:-15.6,0.179036458333333);
\draw[draw=none,fill=steelblue31119180,fill opacity=0.5] (axis cs:-14.4,0.0406901041666667) rectangle (axis cs:-13.6,0.115966796875);
\draw[draw=none,fill=steelblue31119180,fill opacity=0.5] (axis cs:-12.4,0.030517578125) rectangle (axis cs:-11.6,0.0895182291666667);
\draw[draw=none,fill=steelblue31119180,fill opacity=0.5] (axis cs:-10.4,0.0528971354166667) rectangle (axis cs:-9.6,0.154622395833333);
\draw[draw=none,fill=steelblue31119180,fill opacity=0.5] (axis cs:-8.4,0.223795572916667) rectangle (axis cs:-7.6,0.449625651041667);
\draw[draw=none,fill=steelblue31119180,fill opacity=0.5] (axis cs:-6.4,0.510660807291667) rectangle (axis cs:-5.6,0.876871744791667);
\draw[draw=none,fill=steelblue31119180,fill opacity=0.5] (axis cs:-4.4,1.18611653645833) rectangle (axis cs:-3.6,2.17081705729167);
\draw[draw=none,fill=steelblue31119180,fill opacity=0.5] (axis cs:-2.4,3.564453125) rectangle (axis cs:-1.6,5.6884765625);
\draw[draw=none,fill=steelblue31119180,fill opacity=0.5] (axis cs:-0.4,8.09733072916667) rectangle (axis cs:0.4,12.68310546875);
\draw[draw=none,fill=steelblue31119180,fill opacity=0.5] (axis cs:1.6,16.0176595052083) rectangle (axis cs:2.4,22.7030436197917);
\draw[draw=none,fill=steelblue31119180,fill opacity=0.5] (axis cs:3.6,24.4405110677083) rectangle (axis cs:4.4,30.6660970052083);
\draw[draw=none,fill=steelblue31119180,fill opacity=0.5] (axis cs:5.6,30.9163411458333) rectangle (axis cs:6.4,35.1175944010417);
\draw[draw=none,fill=steelblue31119180,fill opacity=0.5] (axis cs:7.6,35.2803548177083) rectangle (axis cs:8.4,38.2527669270833);
\draw[draw=none,fill=steelblue31119180,fill opacity=0.5] (axis cs:9.6,39.8152669270833) rectangle (axis cs:10.4,41.8253580729167);
\draw[draw=none,fill=steelblue31119180,fill opacity=0.5] (axis cs:11.6,43.7744140625) rectangle (axis cs:12.4,45.245361328125);
\draw[draw=none,fill=steelblue31119180,fill opacity=0.5] (axis cs:13.6,46.1527506510417) rectangle (axis cs:14.4,47.8373209635417);
\draw[draw=none,fill=steelblue31119180,fill opacity=0.5] (axis cs:15.6,47.3002115885417) rectangle (axis cs:16.4,49.1414388020833);
\draw[draw=none,fill=steelblue31119180,fill opacity=0.5] (axis cs:17.6,48.1465657552083) rectangle (axis cs:18.4,50.1261393229167);
\draw[draw=none,fill=steelblue31119180,fill opacity=0.5] (axis cs:19.6,48.7223307291667) rectangle (axis cs:20.4,50.4597981770833);
\draw[draw=none,fill=darkorange25512714] (axis cs:-20.4,0.0732421875) rectangle (axis cs:-19.6,0.0732421875);
\addlegendimage{ybar,ybar legend,draw=none,fill=darkorange25512714}
\addlegendentry{Exit 1 Right}

\draw[draw=none,fill=darkorange25512714] (axis cs:-18.4,0.0874837239583333) rectangle (axis cs:-17.6,0.0874837239583333);
\draw[draw=none,fill=darkorange25512714] (axis cs:-16.4,0.179036458333333) rectangle (axis cs:-15.6,0.181070963541667);
\draw[draw=none,fill=darkorange25512714] (axis cs:-14.4,0.115966796875) rectangle (axis cs:-13.6,0.118001302083333);
\draw[draw=none,fill=darkorange25512714] (axis cs:-12.4,0.0895182291666667) rectangle (axis cs:-11.6,0.101725260416667);
\draw[draw=none,fill=darkorange25512714] (axis cs:-10.4,0.154622395833333) rectangle (axis cs:-9.6,0.209554036458333);
\draw[draw=none,fill=darkorange25512714] (axis cs:-8.4,0.449625651041667) rectangle (axis cs:-7.6,0.620524088541667);
\draw[draw=none,fill=darkorange25512714] (axis cs:-6.4,0.876871744791667) rectangle (axis cs:-5.6,1.751708984375);
\draw[draw=none,fill=darkorange25512714] (axis cs:-4.4,2.17081705729167) rectangle (axis cs:-3.6,5.712890625);
\draw[draw=none,fill=darkorange25512714] (axis cs:-2.4,5.6884765625) rectangle (axis cs:-1.6,14.19677734375);
\draw[draw=none,fill=darkorange25512714] (axis cs:-0.4,12.68310546875) rectangle (axis cs:0.4,25.0895182291667);
\draw[draw=none,fill=darkorange25512714] (axis cs:1.6,22.7030436197917) rectangle (axis cs:2.4,34.588623046875);
\draw[draw=none,fill=darkorange25512714] (axis cs:3.6,30.6660970052083) rectangle (axis cs:4.4,45.1171875);
\draw[draw=none,fill=darkorange25512714] (axis cs:5.6,35.1175944010417) rectangle (axis cs:6.4,54.6162923177083);
\draw[draw=none,fill=darkorange25512714] (axis cs:7.6,38.2527669270833) rectangle (axis cs:8.4,60.9354654947917);
\draw[draw=none,fill=darkorange25512714] (axis cs:9.6,41.8253580729167) rectangle (axis cs:10.4,64.7379557291667);
\draw[draw=none,fill=darkorange25512714] (axis cs:11.6,45.245361328125) rectangle (axis cs:12.4,66.9474283854167);
\draw[draw=none,fill=darkorange25512714] (axis cs:13.6,47.8373209635417) rectangle (axis cs:14.4,68.5262044270833);
\draw[draw=none,fill=darkorange25512714] (axis cs:15.6,49.1414388020833) rectangle (axis cs:16.4,69.47021484375);
\draw[draw=none,fill=darkorange25512714] (axis cs:17.6,50.1261393229167) rectangle (axis cs:18.4,70.0113932291667);
\draw[draw=none,fill=darkorange25512714] (axis cs:19.6,50.4597981770833) rectangle (axis cs:20.4,70.611572265625);
\draw[draw=none,fill=darkorange25512714,fill opacity=0.5] (axis cs:-20.4,0.0732421875) rectangle (axis cs:-19.6,0.0874837239583333);
\addlegendimage{ybar,ybar legend,draw=none,fill=darkorange25512714,fill opacity=0.5}
\addlegendentry{Exit 1 Wrong}

\draw[draw=none,fill=darkorange25512714,fill opacity=0.5] (axis cs:-18.4,0.0874837239583333) rectangle (axis cs:-17.6,0.0874837239583333);
\draw[draw=none,fill=darkorange25512714,fill opacity=0.5] (axis cs:-16.4,0.181070963541667) rectangle (axis cs:-15.6,0.18310546875);
\draw[draw=none,fill=darkorange25512714,fill opacity=0.5] (axis cs:-14.4,0.118001302083333) rectangle (axis cs:-13.6,0.124104817708333);
\draw[draw=none,fill=darkorange25512714,fill opacity=0.5] (axis cs:-12.4,0.101725260416667) rectangle (axis cs:-11.6,0.120035807291667);
\draw[draw=none,fill=darkorange25512714,fill opacity=0.5] (axis cs:-10.4,0.209554036458333) rectangle (axis cs:-9.6,0.248209635416667);
\draw[draw=none,fill=darkorange25512714,fill opacity=0.5] (axis cs:-8.4,0.620524088541667) rectangle (axis cs:-7.6,0.740559895833333);
\draw[draw=none,fill=darkorange25512714,fill opacity=0.5] (axis cs:-6.4,1.751708984375) rectangle (axis cs:-5.6,2.24812825520833);
\draw[draw=none,fill=darkorange25512714,fill opacity=0.5] (axis cs:-4.4,5.712890625) rectangle (axis cs:-3.6,7.427978515625);
\draw[draw=none,fill=darkorange25512714,fill opacity=0.5] (axis cs:-2.4,14.19677734375) rectangle (axis cs:-1.6,17.352294921875);
\draw[draw=none,fill=darkorange25512714,fill opacity=0.5] (axis cs:-0.4,25.0895182291667) rectangle (axis cs:0.4,27.5044759114583);
\draw[draw=none,fill=darkorange25512714,fill opacity=0.5] (axis cs:1.6,34.588623046875) rectangle (axis cs:2.4,35.9842936197917);
\draw[draw=none,fill=darkorange25512714,fill opacity=0.5] (axis cs:3.6,45.1171875) rectangle (axis cs:4.4,46.624755859375);
\draw[draw=none,fill=darkorange25512714,fill opacity=0.5] (axis cs:5.6,54.6162923177083) rectangle (axis cs:6.4,56.31103515625);
\draw[draw=none,fill=darkorange25512714,fill opacity=0.5] (axis cs:7.6,60.9354654947917) rectangle (axis cs:8.4,62.860107421875);
\draw[draw=none,fill=darkorange25512714,fill opacity=0.5] (axis cs:9.6,64.7379557291667) rectangle (axis cs:10.4,66.7236328125);
\draw[draw=none,fill=darkorange25512714,fill opacity=0.5] (axis cs:11.6,66.9474283854167) rectangle (axis cs:12.4,69.0409342447917);
\draw[draw=none,fill=darkorange25512714,fill opacity=0.5] (axis cs:13.6,68.5262044270833) rectangle (axis cs:14.4,70.8109537760417);
\draw[draw=none,fill=darkorange25512714,fill opacity=0.5] (axis cs:15.6,69.47021484375) rectangle (axis cs:16.4,71.783447265625);
\draw[draw=none,fill=darkorange25512714,fill opacity=0.5] (axis cs:17.6,70.0113932291667) rectangle (axis cs:18.4,72.4324544270833);
\draw[draw=none,fill=darkorange25512714,fill opacity=0.5] (axis cs:19.6,70.611572265625) rectangle (axis cs:20.4,72.94921875);
\draw[draw=none,fill=forestgreen4416044] (axis cs:-20.4,0.0874837239583333) rectangle (axis cs:-19.6,18.0257161458333);
\addlegendimage{ybar,ybar legend,draw=none,fill=forestgreen4416044}
\addlegendentry{Exit 2 Right}

\draw[draw=none,fill=forestgreen4416044] (axis cs:-18.4,0.0874837239583333) rectangle (axis cs:-17.6,18.267822265625);
\draw[draw=none,fill=forestgreen4416044] (axis cs:-16.4,0.18310546875) rectangle (axis cs:-15.6,20.4752604166667);
\draw[draw=none,fill=forestgreen4416044] (axis cs:-14.4,0.124104817708333) rectangle (axis cs:-13.6,22.1822102864583);
\draw[draw=none,fill=forestgreen4416044] (axis cs:-12.4,0.120035807291667) rectangle (axis cs:-11.6,25.10986328125);
\draw[draw=none,fill=forestgreen4416044] (axis cs:-10.4,0.248209635416667) rectangle (axis cs:-9.6,28.7434895833333);
\draw[draw=none,fill=forestgreen4416044] (axis cs:-8.4,0.740559895833333) rectangle (axis cs:-7.6,31.3313802083333);
\draw[draw=none,fill=forestgreen4416044] (axis cs:-6.4,2.24812825520833) rectangle (axis cs:-5.6,35.6526692708333);
\draw[draw=none,fill=forestgreen4416044] (axis cs:-4.4,7.427978515625) rectangle (axis cs:-3.6,41.8070475260417);
\draw[draw=none,fill=forestgreen4416044] (axis cs:-2.4,17.352294921875) rectangle (axis cs:-1.6,50.286865234375);
\draw[draw=none,fill=forestgreen4416044] (axis cs:-0.4,27.5044759114583) rectangle (axis cs:0.4,60.1603190104167);
\draw[draw=none,fill=forestgreen4416044] (axis cs:1.6,35.9842936197917) rectangle (axis cs:2.4,70.0439453125);
\draw[draw=none,fill=forestgreen4416044] (axis cs:3.6,46.624755859375) rectangle (axis cs:4.4,79.7505696614583);
\draw[draw=none,fill=forestgreen4416044] (axis cs:5.6,56.31103515625) rectangle (axis cs:6.4,84.844970703125);
\draw[draw=none,fill=forestgreen4416044] (axis cs:7.6,62.860107421875) rectangle (axis cs:8.4,87.5834147135417);
\draw[draw=none,fill=forestgreen4416044] (axis cs:9.6,66.7236328125) rectangle (axis cs:10.4,88.5172526041667);
\draw[draw=none,fill=forestgreen4416044] (axis cs:11.6,69.0409342447917) rectangle (axis cs:12.4,89.4327799479167);
\draw[draw=none,fill=forestgreen4416044] (axis cs:13.6,70.8109537760417) rectangle (axis cs:14.4,90.386962890625);
\draw[draw=none,fill=forestgreen4416044] (axis cs:15.6,71.783447265625) rectangle (axis cs:16.4,89.9190266927083);
\draw[draw=none,fill=forestgreen4416044] (axis cs:17.6,72.4324544270833) rectangle (axis cs:18.4,90.7267252604167);
\draw[draw=none,fill=forestgreen4416044] (axis cs:19.6,72.94921875) rectangle (axis cs:20.4,90.997314453125);
\draw[draw=none,fill=forestgreen4416044,fill opacity=0.5] (axis cs:-20.4,18.0257161458333) rectangle (axis cs:-19.6,100);
\addlegendimage{ybar,ybar legend,draw=none,fill=forestgreen4416044,fill opacity=0.5}
\addlegendentry{Exit 2 Wrong}

\draw[draw=none,fill=forestgreen4416044,fill opacity=0.5] (axis cs:-18.4,18.267822265625) rectangle (axis cs:-17.6,100);
\draw[draw=none,fill=forestgreen4416044,fill opacity=0.5] (axis cs:-16.4,20.4752604166667) rectangle (axis cs:-15.6,100);
\draw[draw=none,fill=forestgreen4416044,fill opacity=0.5] (axis cs:-14.4,22.1822102864583) rectangle (axis cs:-13.6,100);
\draw[draw=none,fill=forestgreen4416044,fill opacity=0.5] (axis cs:-12.4,25.10986328125) rectangle (axis cs:-11.6,100);
\draw[draw=none,fill=forestgreen4416044,fill opacity=0.5] (axis cs:-10.4,28.7434895833333) rectangle (axis cs:-9.6,100);
\draw[draw=none,fill=forestgreen4416044,fill opacity=0.5] (axis cs:-8.4,31.3313802083333) rectangle (axis cs:-7.6,100);
\draw[draw=none,fill=forestgreen4416044,fill opacity=0.5] (axis cs:-6.4,35.6526692708333) rectangle (axis cs:-5.6,100);
\draw[draw=none,fill=forestgreen4416044,fill opacity=0.5] (axis cs:-4.4,41.8070475260417) rectangle (axis cs:-3.6,100);
\draw[draw=none,fill=forestgreen4416044,fill opacity=0.5] (axis cs:-2.4,50.286865234375) rectangle (axis cs:-1.6,100);
\draw[draw=none,fill=forestgreen4416044,fill opacity=0.5] (axis cs:-0.4,60.1603190104167) rectangle (axis cs:0.4,100);
\draw[draw=none,fill=forestgreen4416044,fill opacity=0.5] (axis cs:1.6,70.0439453125) rectangle (axis cs:2.4,100);
\draw[draw=none,fill=forestgreen4416044,fill opacity=0.5] (axis cs:3.6,79.7505696614583) rectangle (axis cs:4.4,100);
\draw[draw=none,fill=forestgreen4416044,fill opacity=0.5] (axis cs:5.6,84.844970703125) rectangle (axis cs:6.4,100);
\draw[draw=none,fill=forestgreen4416044,fill opacity=0.5] (axis cs:7.6,87.5834147135417) rectangle (axis cs:8.4,100);
\draw[draw=none,fill=forestgreen4416044,fill opacity=0.5] (axis cs:9.6,88.5172526041667) rectangle (axis cs:10.4,100);
\draw[draw=none,fill=forestgreen4416044,fill opacity=0.5] (axis cs:11.6,89.4327799479167) rectangle (axis cs:12.4,100);
\draw[draw=none,fill=forestgreen4416044,fill opacity=0.5] (axis cs:13.6,90.386962890625) rectangle (axis cs:14.4,100);
\draw[draw=none,fill=forestgreen4416044,fill opacity=0.5] (axis cs:15.6,89.9190266927083) rectangle (axis cs:16.4,100);
\draw[draw=none,fill=forestgreen4416044,fill opacity=0.5] (axis cs:17.6,90.7267252604167) rectangle (axis cs:18.4,100);
\draw[draw=none,fill=forestgreen4416044,fill opacity=0.5] (axis cs:19.6,90.997314453125) rectangle (axis cs:20.4,100);
\end{axis}

\end{tikzpicture}

%% file: 6.conclusions.tex
\section{Conclusion and future work}
This paper introduces and evaluates a novel EE strategy, utilizing width-wise model splitting to alleviate the average computational load. By targeting the computational impact of the initial layer, our proposed architecture achieves a notable reduction in computational load averaging 26.2\% with an average accuracy reduction of 0.7\%.
Notably, the proposed model exhibits expected behavior, demonstrating lower computational loads for higher SNR levels, with reductions of up to 60\%. 

Moreover, this paper proposes a method for dynamically selecting an exit criterion based on the training data by selecting a targeted accuracy and percentage of exiting frames. This method can automate the threshold selection across different instantiations of the proposed model, which can be of special interest to edge computing. 

Future research can focus on optimizing width-wise EE and exploring its applications across diverse domains. In the AMC realm, there is potential for enhancing performance in low-SNR scenarios. Low-SNR frames impose a significant computational load without significantly contributing to accuracy.
Potential optimizations may include sample rejection or redirection to models specialized in handling low-SNR samples. Overall, the findings presented here contribute to the ongoing discourse in deep learning and modulation classification, offering insights and avenues for further exploration in optimizing computational load and model accuracy.

